# *BeppoSAX* observations of the exotic black hole candidate GX 339-4

L.Chiappetti[a], F.Haardt[b], A.Treves[c], D.Ricci[d], A.Santangelo[e], S.Mereghetti[a] and T.Belloni[f]

[a]IFCTR/CNR, Milano, Italy
[b]Dipartimento di Fisica, Universita' di Milano, Italy
[c]Dipartimento di Fisica, Universita' di Milano (Sede di Como), Italy
[d]SAX Science Data Centre, ASI/Telespazio, Roma, Italy
[e]IFCAI/CNR, Palermo, Italy
[f]Astronomical Institute, Amsterdam, The Netherlands

GX 339-4 has been observed by *BeppoSAX* twice in spring 1997 as part of a longer monitoring program. The source was close to the highest levels (50 mCrab) of the extended low state (as measured by the XTE ASM during the last 2 years). Its spectrum was quite hard, similar to the Exosat 1984 off state, but 40 times stronger. The source is detected up to more than 120 keV, enabling the possibility to study its high energy spectrum.

## 1. INTRODUCTION

GX339-4 has been considered a black hole candidate mainly because its X-ray spectrum is the superposition of a soft and a hard component, the former possibly associated with an accretion disk, the latter to a Comptonization hot corona [1, 2 and references therein]. While the two components are rather typical of black hole candidates (e.g. [3] ), the colour/intensity and colour/colour diagrams are reminiscent of Z-sources, which are considered neutron stars with low magnetic field. The optical counterpart is dominated by the X-ray activity as inferred by the very large variability (V= 15-20). A photometric period of 15 h has been proposed [4]. Spectroscopy indicates variability of emission lines velocity, but does not allow an assessment of the mass function. The two possibilities of a black hole or of a neutron star appear both viable.

Comparing with other non transient X-ray binaries, and in particular black hole candidates, the variability is remarkable (see the XTE ASM light curve of the last two years in fig.1; during this time span the source has been in an extended low state, although with variations of a factor 4). A variability of a factor 100 has been reported historically, affecting differently the soft and hard components. In the very high state, on shorter time scale the variability is rather complex with dips and flip-flop intensity variations and quasi periodic oscillations at 6 Hz are well established [2, 5]. Cross-correlation between different X-ray bands suggests the presence of lags, as expected in Comptonization models, possibly occurring in a corona surrounding the accretion disk. Presence of Fe emission and K edge absorption may indicate reprocessing in the accretion disk.

## 2. OBSERVATIONS

GX 339-4 has been observed by *BeppoSAX* as part of a campaign of four 10 ksec pointings (see Table 1). We report here a preliminary analysis of the first two observations (March and April 1997).

It is known from the monitoring of the source conducted by the *XTE* All Sky Monitor (ASM) that the source has been in an extended low state at least since about 2 years. The intensity of the source is much lower than the brightest historically known states and only very recently a brightening to higher level has perhaps started

Table 1
Journal of Observations

| Observation date | Start time (UT) | End time (UT) |
|---|---|---|
| 1997 Mar 17 | 06 :59 | 12 :56 |
| 1997 Apr 06 | 10 :13 | 15 :55 |
| 1997 Sep 06 | 05 :22 | 12 :54 |
| 1997 Oct 01-02 | 17 :17 | 05 :48 |

In Fig. 1 we report the *XTE* ASM light curve [6], with the indication of the epoch of the four *BeppoSAX* observations. The two *BeppoSAX* observations examined here are close to the highest level attained during the ASM lifetime until the recent brightening.

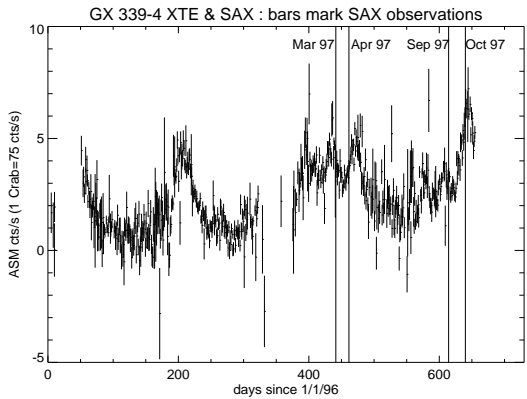

Figure. 1. The XTE ASM one-day average light curve with the indication of the epoch of our *BeppoSAX* observations. This paper describes only the first two *BeppoSAX* observations.

## 3. DATA REDUCTION

The source has been detected by all four Narrow Field Instruments onboard *BeppoSA*X at a level of the rough order of 50 mCrab. We report the live exposure time and count rates in selected energy bands in Table 2.

The MECS, HPGSPC and PDS data have been reduced with the XAS software [7], while the LECS data have been reduced using the SDC standard analysis based on SAXDAS.

The LECS data have been extracted from an 8.1 arcmin region, and using the standard background. The LECS has been operated only in shadow as customary, hence its exposure times are lower.

The MECS data have been extracted from a 13 arcmin region. A preliminary background subtraction has been done using a surrounding 19 arcmin ring. This is known to underestimate the background [8], which is however two orders of magnitude fainter than the source. A further more detailed analysis to be done will involve the rejection or recovery of some intervals where the attitude was not perfect (less than 2 star trackers in operation).

In the March observation the HPGSPC has been operated using the old collimator law (in which on-source (aligned) stay times of 96 s are interleaved with offset stays in both positive and negative directions). In the April observation the new collimator law has been used (in which offset stays are collected only from negative directions). The standard data extraction with Burst Length thresholds 85-115 has been applied. While the standard background subtraction is fine for the March pointing, for the April pointing it shows some residual features, and an alternate method has been used (subtracting aligned stays taken during source occultation).

The PDS data have been reduced in the standard way, which implies the accumulation of separate spectra for the 4 PDS units during the intervals when the collimators are aligned or offset in the positive and negative direction. There are two units under each collimator, and the motions of the two collimators are anticorrelated so that one half is always looking at the source. The background (determined from the offset intervals) is sub-tracted by the aligned spectra for each unit, the four net spectra are gain-equalized, rebinned and finally combined into a single one. In particular we note that signal from the source can be detected at a 5 σ level in the PDS up to 130 keV.

Table 2
Summary of count rates

| Instrument | Band (keV) | 17 March 1997 | | 06 Apr 1997 | |
|---|---|---|---|---|---|
| | | exposure (s) | cts/s | exposure (s) | cts/s |
| LECS | 2-4 | 4979 | 6.33±0.03 | 7445 | 5.48±0.03 |
| MECS M1 | 2-10.5 | 10794 | 5.29±0.02 | 10793 | 4.67±0.02 |
| MECS M2 | " | 10795 | 5.94±0.02 | 12183 | 5.27±0.02 |
| MECS M3 | " | 10793 | 6.12±0.02 | 11276 | 5.40±0.02 |
| HPGSPC | 7-40 | 4273 | 19.5±0.2 | 5587 | 18.8±0.2 |
| PDS | 20-140 | 4701+4761 | 18.7±+0.1 | 5860+5889 | 17.1±0.1 |

The source shows no striking evidence of variability *within* each observations (we plan to do a detailed temporal analysis in the future). There is also no large variation in intensity between the March and April observations.

## 4. PRELIMINARY RESULTS

The spectral fits reported below are to be considered preliminary, also at the light of the current knowledge of the intercalibration of the various instruments.

The MECS data alone can be fitted by a single PL (photon index 1.6) in the range 2-10 keV. Of course this instrument is not particularly apt to determine the hydrogen column density.

Combining MECS and LECS (0.2-4 keV) data, we are able to determine $N_H = 5.1 \times 10^{21}$ cm$^{-2}$ as expected. In addition there is a clear evidence of the presence of a soft excess. We are so far unable to discriminate between a blackbody and a second power law component (e.g. broken power law)

The PDS data alone show a clear evidence of a spectral break about 80 keV These data cannot be fitted by a single power law.

Introducing an exponential cutoff its energy is found to be 82 keV, but the power law is extremely flat (photon index 1.2).

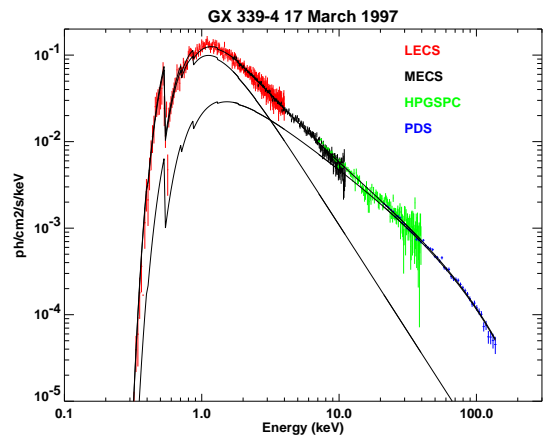

Figure. 2. Deconvolved photon spectrum of GX339-4 as observed in March 1997 by the four *BeppoSAX* NFIs. Both components of the model described in the text are shown. MECS data are for detector M1.

A combined fit of PDS with the two imaging instruments (LECS+MECS), while grossly confirming the power law shape in the 2-50 kev range (and obviously the low energy excess), and the presence of the high energy cutoff, encounters some difficulties with positive residuals in the 20-70 keV range (additional component ?), and negative residuals in the high energy band.

The insertion of the HPGSPC data to bridge the gap between MECS and PDS gives a somewhat different scenario. A joint HPGSPC-PDS fit allows to fix the cross-normalization at 1 :0.8. After that one obtains a not unreasonable fit ($\chi^2$=526.2/427 DoF) to all four instruments with two components : e.g. for the March spectra, a "hard" power law with photon index 1.13±0.05, 1 keV normalization of 0.07±0.01 photon/cm$^2$/s/keV (and a cutoff at 80.5±6.5 keV) and a "soft" power law with spectral index 2.47±0.12 and 1 keV normalization of 0.33±0.01 photon/cm$^2$/s/keV. Errors quoted are 1 σ. The PDS and LECS data have been normalized with respect to the other two instruments by a factor 0.8 and 0.87 respectively

Since all this points out anyhow to a rather complicated spectrum, we plan to undertake further more detailed analysis.

The spectral parameters are extremely similar between the March and April observations (the April spectrum being scaled down in intensity by 90%), therefore we report in Fig. 2 just one of the 2 spectra.

We note incidentally that the power law slope found with the MECS alone is quite similar to the one observed in a fainter low state in 1985 by *Exosat*, with a similar energy range. This source was otherwise known to be one of the softest sources when in its high state. The presence of a soft component is resemblant of the behaviour of the well known black hole candidate Cyg X-1 in its hard state, suggesting once more that also GX339-4 may belong to the same class.

# ACKNOWLEDGMENTS

We gratefully acknowledge browsing the *XTE* light curves provided on line by MIT CSR.

We warmly thank the staff of the *BeppoSAX* Science Data Centre for the planning and execution of our observations.

The present research has been accomplished despite the actions by the Italian governmental bureaucracy (and in particular by ARAN) to degrade the work conditions of research staff of Public Research Organizations (PRO), like CNR, and of their attempt to introduce a radical disruption of the unity of research between PRO and Universities.